\documentclass[aps,pra,reprint,superscriptaddress,amsfonts,amsmath,amssymb,notitlepage]{revtex4-1}

\usepackage{graphicx,adjustbox}
\usepackage[caption=false]{subfig}
\usepackage{bbm}
\usepackage{pgfplots,tikz}

\usetikzlibrary{shapes,backgrounds,fit,decorations.pathreplacing,calc,external}
\makeatletter
\tikzset{circle split part fill/.style  args={#1,#2}{%
 alias=tmp@name, 
  postaction={%
    insert path={
     \pgfextra{%
     \pgfpointdiff{\pgfpointanchor{\pgf@node@name}{center}}%
                  {\pgfpointanchor{\pgf@node@name}{east}}%
     \pgfmathsetmacro\insiderad{\pgf@x}
      \fill[#1] (\pgf@node@name.base) ([xshift=-\pgflinewidth]\pgf@node@name.east) arc
                          (0:180:\insiderad-\pgflinewidth)--cycle;
      \fill[#2] (\pgf@node@name.base) ([xshift=\pgflinewidth]\pgf@node@name.west)  arc
                           (180:360:\insiderad-\pgflinewidth)--cycle;            
         }}}}}  
 \makeatother

\begin{document}


\title{Efficient state initialization by a quantum spectral filtering algorithm}


\author{Fran\c{c}ois Fillion-Gourdeau}
\email{francois.fillion@emt.inrs.ca}
\affiliation{Universit\'{e} du Qu\'{e}bec, INRS-\'{E}nergie, Mat\'{e}riaux et T\'{e}l\'{e}communications, Varennes, Canada J3X 1S2}
\affiliation{Institute for Quantum Computing, University of Waterloo, Waterloo,
Ontario N2L 3G1, Canada}

\author{Steve MacLean}
\email{steve.maclean@emt.inrs.ca}
\affiliation{Universit\'{e} du Qu\'{e}bec, INRS-\'{E}nergie, Mat\'{e}riaux et T\'{e}l\'{e}communications, Varennes, Canada J3X 1S2}
\affiliation{Institute for Quantum Computing, University of Waterloo, Waterloo,
Ontario N2L 3G1, Canada}

\author{Raymond Laflamme}
\email{laflamme@iqc.uwaterloo.ca}
\affiliation{Institute for Quantum Computing, University of Waterloo, Waterloo,
Ontario N2L 3G1, Canada}
\affiliation{Department of Physics and Astronomy, University of Waterloo, Waterloo,
Ontario N2L 3G1, Canada}
\affiliation{Perimeter Institute for Theoretical Physics, Waterloo, Ontario N2L 2Y5, Canada}
\affiliation{Canadian Institute for Advanced Research, Toronto, Ontario M5G 1Z8, Canada}


\date{\today}

\begin{abstract}
An algorithm that initializes a quantum register to a state with a specified energy range is given,  corresponding to a quantum implementation of the celebrated Feit-Fleck method. This is performed by introducing a nondeterministic quantum implementation of a standard spectral filtering procedure combined with an apodization technique, allowing for accurate state initialization. It is shown that the implementation requires only two ancilla qubits. A lower bound for the total probability of success of this algorithm is derived, showing that this scheme can be realized using a finite, relatively low number of trials. Assuming the time evolution can be performed efficiently and using a trial state polynomially close to the desired states, it is demonstrated that the number of operations required scales polynomially with the number of qubits. Tradeoffs between accuracy and performance are demonstrated in a simple example: the harmonic oscillator.  This algorithm would be useful for the initialization phase of the simulation of quantum systems on digital quantum computers. 
\end{abstract}

\pacs{}

\maketitle

\section{Introduction}

Quantum simulation is one of the most important applications of quantum computing because the time evolution of many physical systems can be realized with a number of quantum gates scaling logarithmically with the system size, opening up the possibility of making calculations in regimes inaccessible on classical computers \cite{feynman1982simulating,SLoyd,RevModPhys.86.153}. Many quantum algorithms have been developed to simulate efficiently the dynamics of quantum systems, such as single particle non-relativistic systems \cite{OPPROP:PR877,Zalka08011998,Strini2008}, many-body systems \cite{boghosian1998simulating}, fermionic systems \cite{PhysRevLett.79.2586,PhysRevA.64.022319} and many others \cite{yung2014}. 

Although many efficient algorithms have been discovered for the simulation of the time evolution of quantum systems, the initialization of the quantum register to a desired initial state remains a challenge. Ideally, this state would be physically relevant while being efficiently implemented on the quantum computer. For general initial states, this cannot be achieved because it involves the execution of diagonal unitary gates requiring $O(2^{n+1})$ operations, where $n$ is the number of qubits \cite{PhysRevLett.91.027902,PhysRevA.71.052330,bullock2004asymptotically}. This can be improved by using Walsh basis functions techniques and by approximating the unitary operation \cite{1367-2630-16-3-033040}. Therefore, these ``brute-force'' techniques entail an exponential number of gates and thus, deteriorate the performance of any quantum simulation algorithm.

A quantum algorithm for the initialization of the quantum register has also been formulated for the simulation of real space non-relativistic quantum mechanics \cite{Zalka08011998,kaye2004quantum,grover2002creating}. The initial state is constructed incrementally by adding qubit contributions while conserving the probability distribution. This method also requires diagonal unitaries for general states, but may be efficient for a certain class of function. 

Another popular approach for the initialization is the phase-approximation method pioneered by Abrams and Lloyd \cite{PhysRevLett.83.5162}. The Abrams and Lloyd technique (ALT) yields eigenvalues and eigenstates of time-independent Hamiltonian operators by evolving a trial state in time and by separating its spectral component using a quantum Fourier transform on an ancilla qubit register. If the trial state is polynomially close to the eigenstate, i.e. if the overlap between the trial state and the eigenstate is bounded by a polynomial function of the problem size, success can be achieved in polynomial time because the success probability of the algorithm is proportional to this overlap. Conversely, this probability can be exponentially small if the eigenstate component in the trial state is unimportant compared to other modes, although this can be improved by means of the adiabatic-state-preparation algorithm \cite{Aspuru-Guzik1704}. The ALT has been employed in the context of chemical physics to assess the possibility of determining eigenstates of molecular systems on quantum computers \cite{Aspuru-Guzik1704,Kassal02122008}. Despite its very interesting properties and success, this scheme usually requires a relatively large number of ancilla qubits to properly resolve the eigenenergies and eigenstates.

In this article, we present an alternative approach to the initialization problem which bears some resemblance to the ALT but differs on two main aspects:  (1) an apodization technique is introduced to improve the accuracy of the generated states, (2) a non-unitary operation is introduced to decrease the number of required ancilla qubits.  

The goal of our algorithm is to set the value of the amplitude of a quantum register with the resulting state $|\Psi_{\rho}\rangle$ having spectral components within a specified energy range.
 Our algorithm takes advantage of the efficient time evolution to initialize the register by a spectral filtering technique: a trial function is evolved in time and filtered to keep only the desired spectral components. It can be seen as a quantum generalization of the  Feit-Fleck spectral method, originally developed to evaluate eigenenergies and eigenstates of the Schr\"{o}dinger equation in a static potential by solving the time-dependent dynamics \cite{Feit1982412}. The latter has been used successfully in a wide range of applications on classical computers \cite{feit1983,Mocken2004558,FillionGourdeau2014559}.

The quantum algorithm presented in the following has some interesting properties. First, it requires only two ancilla qubits. Second, assuming that the overlap between the trial state and the desired state is not negligible, it yields the desired state with a relatively high probability (see Eq. \eqref{eq:prob_success}) and it can be implemented efficiently. Third, it can accommodate states with many spectral components, such as wave packets. However, the weight of each component is connected to their value in the trial function, prior to filtering. Finally, the implementation naturally allows for an apodization function which improves the filtering and the accuracy of the desired states. Hereinafter, we shall assess each of these properties. 

This article is separated as follows. Section \ref{sec:quantum_sim} reviews some basic facts on quantum simulations. In Sec. \ref{sec:qu_spec_filt}, the quantum implementation of the spectral filtering method is presented. Sec. \ref{sec:complexity_an} is devoted to the complexity analysis and the resource requirements for the algorithm. A simple example where the ground state of the harmonic oscillator is generated from the filtering method is considered in Sec. \ref{sec:harm_osci}. Finally, the conclusion is found in Sec. \ref{sec:conclusion}.

\section{Quantum simulations}
\label{sec:quantum_sim}

In this section, a brief review of quantum simulations on digital quantum computers is given. The main goal here is to present the notation and to describe the objective of our initialization algorithm.

To simulate a quantum system on a quantum computer, one has to find a set of quantum gates that approximates the exact evolution operator
\begin{align}
\label{eq:evol_op}
\hat{U}(0,T) = \exp\left[-iT \hat{H}  \right],
\end{align}
where $T$ is the final time and $\hat{H}$ is the Hamiltonian describing the physical system under investigation. For the more general case of a time-dependent Hamiltonian, this would become a time-ordered exponential \footnote{For time-dependent Hamiltonian, the evolution operator is $\hat{U}(0,T) = \mathcal{T}\exp\left[-i \int_{0}^{T}dt \hat{H}(t)  \right]$, where $\mathcal{T}$ is the time-ordering operator.}.

The evolution operator usually cannot be evaluated exactly and one has to resort to some approximation scheme. One popular method is based on the Trotterization of the exponential where the total time is split into $N_{t}$ small intervals of duration $\Delta t = T/N_{t}$, as
%
$\hat{U}(0,T) = U_{1} \cdots U_{N_{t}}$.
%
Then, each evolution operator $U_{i}:=\hat{U}((i-1)\Delta t, i \Delta t)$ is approximated by the Trotter-Suzuki exponential product $\hat{U}_{\mathrm{approx}}$ with an accuracy $O(\Delta t^{q+1})$, where $q \in \mathbb{N}^{+}$ gives the order of the approximation. The order can be improved to arbitrarily large value \cite{suzuki1993general}. The resulting scheme is a product of unitary transformations which can be simulated on a quantum computer using quantum gates \cite{nielsen2010quantum}. These gates are applied on a quantum register made of $n$ qubits where the state of the register is given by a $2^{n}$-dimensional vector in the Hilbert space $\mathcal{H}_{n} = \bigotimes_{i=1}^{n} \mathcal{H}_{1}$, where $\mathcal{H}_{1}$ is the Hilbert space of one qubit. The state of the register is then expressed as
\begin{align}
|\psi_{n} \rangle 
&=   \sum_{k = 1}^{2^{n}} 
\alpha_{k} 
 | k \rangle,
\end{align}
where $(\alpha_{k})_{k=1,\cdots,2^{n}}$ are complex amplitudes and $| k \rangle$ represents the tensor product of $n$ qubit states, the sum is carried over all binary strings. Similarly, the state of the physical system to be simulated is described by a vector in a Hilbert space $\mathcal{H}$. Such vectors can be approximately written as
\begin{align}
|\Psi(t) \rangle \approx \sum_{k=1}^{2^{n}} b_{k}(t) |\phi_{k} \rangle,
\end{align}     
where $b_{k}(t)$ are time-dependent coefficients and $|\phi_{k} \rangle$ are orthonormal basis vectors that span a Hilbert space $\mathcal{H}_{\rm approx}$ with dimension $2^{n}$, matching the dimension of the quantum register. It is assumed that vectors in the Hilbert space $\mathcal{H}_{\rm approx}   \subseteq \mathcal{H}$ are accurate approximations of vectors in the full Hilbert space of the physical system $\mathcal{H}$. This occurs when $n$ is large enough, assuming the convergence of the discrete Hilbert space to some region in the Hilbert space of the physical system. Then, the coefficients $b_{k}$ can be directly mapped to the coefficients $\alpha_{k}$ as $b_{k} \mapsto \alpha_{k}$, allowing the quantum computer to store the discretized state of the physical system under consideration. At the same time, the approximate evolution operator expressed in the physical basis as $(U_{\rm approx})_{jk} = \langle \phi_{j}|\hat{U}_{\rm approx}|\phi_{k} \rangle$ is mapped to quantum gates. Applying these operations on the quantum register, the coefficients $\alpha_{k}$ store the approximate time evolution of the state. When the number of gates obtained from the mapping scales like $\mathrm{poly}(n)$ for a given constant precision $\epsilon \ll 1$, we shall say that the resulting algorithm is efficient. This whole strategy is the essence of quantum simulations on digital quantum computers.

The main goal of the algorithm presented in this article is the initialization of the quantum register $|\psi_{n}\rangle$ to an initial state relevant for quantum simulations. Of course, this step must occur before the actual time evolution is implemented. The filtering technique presented in the next section sets the value of the coefficients $\alpha_{k}$, ensuring that the register represents a discretized eigenstate of a given time-independent Hamiltonian. These initial eigenstates are typical in numerical simulations of quantum systems on classical computers in atomic, molecular and optics physics, for example \cite{doi:10.1021/j100319a003,tannor2007introduction}.

\section{Quantum spectral filtering method}
\label{sec:qu_spec_filt}

The basic ingredient of the spectral filtering method is the observation that an exact state $|\varphi_{\rho}\rangle$ with a spectral content $\rho$ can be approximated from
\begin{align}
\label{eq:eigen_state}
|\varphi_{\rho}\rangle \approx |\Psi_{\rho}\rangle = \frac{1}{T} \int_{0}^{T} dt w_{\rho}(t) e^{iE_{\rho}t} |\Psi_{\rm trial}(t)\rangle,
\end{align}
where $T$ is is final time of the calculation, $w_{\rho}(t)$ is the window (apodization) function, $E_{\rho}$ is the central frequency and $|\Psi_{\rm trial}(t)\rangle$ is a time-dependent state initialized to an arbitrary trial value. This last formula is a simple consequence of the trial state eigendecomposition, as demonstrated in Appendix \ref{app:filtering}. 

Here, $w_{\rho}(t)$ is the window function that selects the wanted spectral component and that accounts for the fact that a finite time evolution is performed. It is normalized such that $w_{\rho}(t)\in [0,1]$ and $\max_{t\in [0,T]} w_{\rho}(t) =1$. For an infinite time evolution, the spectrum would become a weighted Dirac comb. On the other hand, if the window function is rectangular ($w_{\rho}(t) = 1$ for $t \in [0,T]$), the spectral peaks will be given by a sequence of $\mathrm{sinc}(E_{\rho})$ functions which have large spectral leakage, i.e. they have large spectral components outside the central frequency. This phenomenon occurs in the standard ALT. A window function which vanishes smoothly close to $t=0$ and $t=T$ reduces spectral leakage by improving the suppression of unwanted modes. As shown in Eq. \ref{eq:spec_coeff} of Appendix \ref{app:filtering}, the magnitude of unwanted modes is suppressed by the line shape centered on energy $E_{\rho}$, given by
\begin{align}
\label{eq:line_shape}
L_{\rho}(E_{\rho}-E_{m}) = \frac{1}{T} \int_{0}^{T}dt e^{i(E_{\rho}-E_{m})t} w_{\rho}(t),
\end{align}
where $E_{m}$ is the (discrete) energy of the state (see Appendix \ref{app:filtering}). 
The strength of the suppression is dictated by the choice of the window function. Having a large suppression factor is important when the overlap of the trial state and the desired state is small as $ A:=|\langle \varphi_{\rho}|\Psi_{\rm trial}(0)\rangle| \ll 1 $. In this case, the spectral coefficient of the desired state $|a_{\rho}|^{2}$ in the trial state also has a small contribution. For this mode to dominate after the filtering procedure, the suppression factor has to be at least as high as the ratio between $|a_{\rho}|^{2}$ and the maximum spectral coefficient of other modes. Then, a window function with a high suppression factor could, in principle, filter the unwanted mode successfully and in turn, this would increase the accuracy of the eigenstate estimation. However, as demonstrated in the following, $A$ is also proportional to the total probability of success of the quantum algorithm.

The algorithm described below requires the energy $E_{\rho}$ of the desired state to be known beforehand. For some cases, this can be obtained from analytical or classical computational methods. There also exist efficient quantum computation approaches whereby the spectrum is calculated semi-classically from the autocorrelation function obtained by performing a measurement  of $\langle \sigma_{x,y}\rangle$, where $\sigma_{x,y}$ are Pauli matrices, on one added ancilla qubit \cite{PhysRevLett.81.5672,PhysRevA.65.042323}. An explicit implementation of this approach relevant to the initialization is given in Appendix \ref{app:en_spec}.

Then, the determination of the eigenstate proceeds in two stages: (1) choosing a trial state and (2) evolving this trial state in time while evaluating Eq. \eqref{eq:eigen_state}. These steps can be implemented on a quantum computer by supplementing the quantum register with an additional qubit $|c \rangle$. The Hilbert space of this qubit serves to label whether the quantum register $| \psi_{n} \rangle $ stores the trial function $|\Psi_{\rm trial}(t)\rangle$  or the constructed initial state $|\Psi_{\rho}\rangle$. The quantum register should be initialized in the state $|\psi_{n+1}\rangle:=|c\rangle \otimes| \psi_{n} \rangle =|0\rangle \otimes |00 \cdots 0\rangle$.

\begin{figure*}
\includegraphics[width=0.7\textwidth]{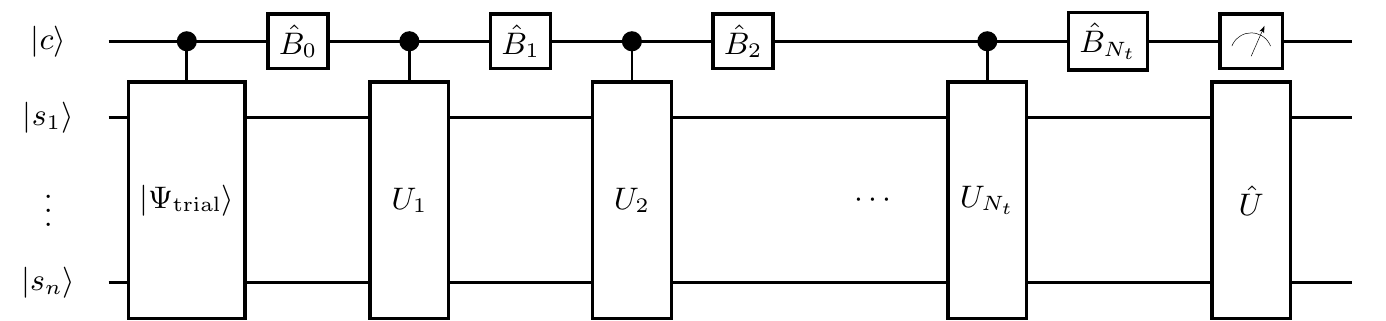}
\caption{Circuit diagram for the quantum implementation of the spectral filtering method. The gate $U_{i}$ advances the trial solution by $\Delta t$. The gate $B_{i}$ is a non-unitary operation. It is defined in Eq. \eqref{eq:gate_B} and its quantum circuit is displayed in Fig. \eqref{fig:nonunitary}. The gate $|\Psi_{\rm trial}\rangle$ initializes the quantum register to an arbitrary state. The projective measurement operator implements the projective measurement $|1\rangle \langle 1|$ that collapses the register to the wanted wave function. Finally, the gate $\hat{U}$ performs the quantum simulation by evolving in time the initial state constructed by the filtering algorithm. }
\label{fig:init_ff}
\end{figure*}

In the first step, the trial function is  constructed from controlled gates, as displayed in Fig. \ref{fig:init_ff}. At this point, the quantum register is in the state $| \psi_{n+1} \rangle =|0\rangle \otimes |\Psi_{\mathrm{trial}}(0)\rangle$, implying that the norm of the trial state is $\langle \Psi_{\mathrm{trial}}(0)|\Psi_{\mathrm{trial}}(0)\rangle =1$. Moreover, we assume that this trial state can be constructed using a number of gates scaling like $\mathrm{poly}(n)$. This is possible in principle because it is an arbitrary state where the coefficient $\alpha_{k}$ can take any value. One possible way to implement the trial state is the utilization of the technique described in Refs. \cite{Zalka08011998,kaye2004quantum,grover2002creating}, which allows for the efficient initialization of a certain class of function. Another approach is given in Ref. \cite{PhysRevLett.91.257902} where an approximation of the eigenstate is obtained by an efficient grid refinement. The trial state can also be a thermal random state as in the \textit{deterministic quantum computation with one quantum bit} (DQC1) model of computation \cite{PhysRevLett.81.5672}. The minimal requirement is that the overlap with the desired spectral components should not be exponentially small, as discussed further in the next section. 


For the second step, we use the fact that the filtering of the trial function given in Eq. \eqref{eq:eigen_state} can be approximated by a quadrature formula of the form
\begin{align}
\label{eq:eigen_state_quad}
|\Psi_{\rho}\rangle &=  \sum_{i=0}^{N_{t}} B_{i}|\Psi_{\rm trial}(t_{i})\rangle + O(\Delta t^{q}) ,\\
B_{i} &:= \frac{u_{i} w_{\rho}(t_{i})e^{iE_{\rho}t_{i}}}{N_{t}} ,
%
\end{align}
where $N_{t}$ is the number of time steps, $t_{i} := i\Delta t$ is the time where the integrand is evaluated and $(u_{i})_{i=0,\cdots,N_{t}}$ are coefficients defined by the quadrature rule \footnote{For example, for the trapezoidal rule, we have $u_{0} = u_{N_{t}} = 1/2$ and $(u_{i})_{i=1,\cdots,N_{t}-1}=1$.}: they are chosen such that the order of accuracy is at least $(\Delta t)^{q} \approx \epsilon$ to ensure that this numerical error does not dominate over the error attributable to the time evolution approximation. The sum in Eq. \eqref{eq:eigen_state_quad} is then computed by alternating a gate $\hat{B}_{i}$ with a controlled gate c-$U_{i+1}$. The latter evolves the trial function by one time step while $\hat{B}_{i}$ actually performs the sum. The quantum circuit associated to this algorithm is displayed in Fig. \ref{fig:init_ff}. 

As mentioned earlier, it is assumed throughout that the time evolution implemented by the gates c-$U_{i+1}$ is efficient, i.e. that it yields an accurate approximation of Eq. \eqref{eq:evol_op} in $\mathrm{poly}(n) $ number of gates. This property of the time-dependent solver is very important for the global efficiency of the initialization algorithm.

The operator $\hat{B}_{i}$ is a non-unitary operator applied on the added qubit $|c \rangle$ and is given in the computational basis by
\begin{align}
\label{eq:gate_B}
\hat{B}_{i} :=  \frac{1}{ \sqrt{1+ \frac{|B_{i}|^{2}}{2} + |B_{i}|\sqrt{1+\frac{|B_{i}|^{2}}{4}}}}
\begin{bmatrix}
1 & 0 \\
B_{i} & 1
\end{bmatrix},
\end{align} 
where the prefactor changes the normalization and guarantees that the operator $\hat{B}_{i}$ can be \textit{literally realized} \cite{Blass2015}. It can be checked that after $N_{t}$ iterations, the quantum register will be in the state
\begin{align}
\label{eq:iterations}
|c\rangle \otimes |\psi_{n} \rangle& = \frac{|0\rangle \otimes |\Psi_{\mathrm{trial}}(T)\rangle +  |1\rangle \otimes |\Psi_{\rho}\rangle}
{N}, \\
N &:= \sqrt{\langle \Psi_{\mathrm{trial}}(T)|\Psi_{\mathrm{trial}}(T)\rangle + \langle \Psi_{\rho} |\Psi_{\rho}\rangle}, \\
&= \sqrt{1 + \langle \Psi_{\rho} |\Psi_{\rho}\rangle},
\end{align}
where the last equality is obtained because unitary operations are used to evolve the trial state and thus, do not change its norm.

Performing a projective measurement $|1\rangle \langle 1|$ (see Fig. \ref{fig:init_ff}), one obtains that $|c\rangle \otimes |\psi_{n} \rangle \mapsto |1\rangle \otimes |\Psi_{\rho} \rangle$. This state
can then be used as an initial condition for the time-dependent simulation of a physical system. The probability of success of the projective measurement is related to the probability to be in the eigenstate, i.e. $P_{\rho} := \langle \Psi_{\rho} | \Psi_{\rho}\rangle/N^{2}$. 


The main challenge of this strategy is the implementation of the non-unitary operation $\hat{B}_{i}$. Such operations and their quantum gate decomposition have already been considered \cite{Terashima2005,mezzacapo2015quantum,Blass2015,Childs:2012:HSU:2481569.2481570,PhysRevLett.114.090502} and a similar strategy is utilized here. First, the matrix $\hat{B}_{i}$ is re-written as
%
$\hat{B}_{i} = U_{i} \Sigma_{i} V^{\dagger}_{i}$, 
%
obtained from the usual singular value decomposition. This yields two unitary matrices $U_{i}, V^{\dagger}_{i}$ and the diagonal matrix $\Sigma_{i} = \mathrm{diag}(1,s_{i})$. The matrices $U_{i}, V^{\dagger}_{i}$ can be evaluated explicitly and are given in Appendix \ref{app:svd}. The second singular value is
\begin{align}
s_{i} &= \sqrt{ \frac{1+ \frac{|B_{i}|^{2}}{2} - |B_{i}|\sqrt{1+\frac{|B_{i}|^{2}}{4}}}{1+ \frac{|B_{i}|^{2}}{2} + |B_{i}|\sqrt{1+\frac{|B_{i}|^{2}}{4}}}} .
\end{align}
The value of the prefactor in Eq. \eqref{eq:gate_B} was chosen such that the first singular value is ``1'' while the second obey $|s_{i}| \leq 1$, in accordance with the exact realization theorem \cite{Blass2015}. Then, the probability interpretation is preserved and the operator $\Sigma_{i}$ can be literally realized with one ancilla qubit, a unitary transformation and a projective measurement. This procedure is now detailed.

\begin{figure}
\includegraphics[width=0.5\textwidth]{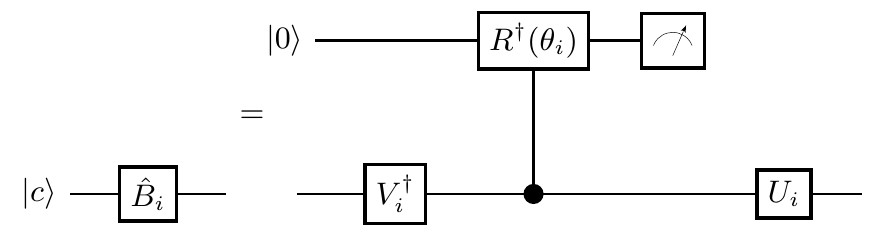}
\caption{Circuit diagram for the implementation of the nonunitary operation. The upper qubit is an ancilla qubit prepared in the state $|0\rangle$. The measurement operator implements the projective measurement $|0\rangle \langle 0|$. If the measurement yields the state $|1\rangle$, the calculation has to be redone from the beginning.}
\label{fig:nonunitary}
\end{figure}

Introducing another ancilla in the state $|0\rangle$, the matrix $\Sigma_{i}$ can be decomposed into the following transformations (up to a normalization):
\begin{align}
& |0\rangle \otimes V_{i}^{\dagger} \left(|0\rangle \otimes |\Psi_{\mathrm{trial}}(i\Delta t)\rangle + |1\rangle \otimes |\Psi^{(i-1)}_{\rho}\rangle \right)= \nonumber \\
& |0\rangle \otimes \left(|0\rangle \otimes |\Psi_{0}^{(i)} \rangle + |1\rangle \otimes |\Psi_{1}^{(i)}\rangle \right) \\
\label{eq:map_nonunit}
&\mapsto   |00 \rangle  \otimes |\Psi_{0}^{(i)}\rangle + \left( s_{i} |01\rangle + \sqrt{1-s_{i}^{2}} |11\rangle \right) \otimes |\Psi_{1}^{(i)}\rangle \\
\label{eq:map_meas}
&\mapsto |0\rangle \otimes \left( |0 \rangle \otimes |\Psi_{0}^{(i)} \rangle + s_{i} |1\rangle \otimes |\Psi_{1}^{(i)}\rangle  \right) \\
\label{eq:map_last}
&\mapsto |0\rangle \otimes U_{i} \left( |0 \rangle \otimes |\Psi_{0}^{(i)} \rangle + s_{i} |1\rangle \otimes |\Psi_{1}^{(i)}\rangle  \right) = \nonumber \\
&\quad \; \; |0\rangle \otimes \left(|0\rangle \otimes |\Psi_{\mathrm{trial}}(i\Delta t)\rangle + |1\rangle \otimes |\Psi^{(i)}_{\rho}\rangle \right),
\end{align}
where $|\Psi^{(j)}_{\rho}\rangle$ is the partial sum (for $i=0,\cdots,j$) in Eq. \eqref{eq:eigen_state_quad} and where $|\Psi_{0,1}^{(i)}\rangle$ are linear combinations of $|\Psi_{\mathrm{trial}}(i\Delta t)\rangle$ and $|\Psi^{(i-1)}_{\rho}\rangle$ obtained by applying $V_{i}^{\dagger}$ on the qubit $|c\rangle$.
The first mapping (Eq. \eqref{eq:map_nonunit}) is performed by a unitary operation in the subspace of the ancilla and the $|c\rangle$ qubits corresponding to a controlled inverse rotation where the rotation angle $\theta_{i}$ is $\cos(\theta_{i})= s_{i}$, controlled by the qubit $|c\rangle$. In turn, Eq. \eqref{eq:map_meas} is obtained by a non-unitary projective measurement $|0\rangle \langle 0|$ on the second ancilla qubit while Eq. \eqref{eq:map_last} is achieved by applying the operator $U_{i}$ on the first ancilla qubit $|c \rangle$. The corresponding circuit diagram is displayed in Fig. \ref{fig:nonunitary}. Going from Eq. \eqref{eq:map_nonunit} to Eq. \eqref{eq:map_meas}, the normalization changes from $N_{i-1}^{2}$ to $N_{i}^{2}$, where
\begin{align}
N_{i} := \sqrt{ 1 + \langle \Psi^{(i)}_{\rho}|\Psi^{(i)}_{\rho}\rangle }.
\end{align}


This algorithm is nondeterministic: every time the projective measurement is performed, the value of the ancilla qubit is verified. If it is in the state $|0\rangle$, this is a ``success'' and the rest of the algorithm can follow because the non-unitary operation has been performed properly. Otherwise, if it is in the state $|1 \rangle$, this is a ``failure'' and the initialization phase has to be reworked from the beginning. Each measurement has a definite success and failure probability. From Eq. \eqref{eq:map_nonunit}, the probability of failure of one projective measurement at step $i=0,\cdots,N_{t}$ is given by
\begin{align}
\label{eq:p_failure}
p_{\mathrm{failure}}^{(i)} = \frac{\langle \Psi_{1}^{(i)}|\Psi_{1}^{(i)} \rangle  (1-s_{i}^{2})}{N_{i-1}^{2}}.
\end{align}
%
%
%
The failure probability is maximized when $p_{1}^{(i)}:=\langle \Psi_{1}^{(i)}|\Psi_{1}^{(i)} \rangle/N_{i-1}$ is maximal. It is demonstrated in Appendix \ref{app:bound_p} that the maximization over the norm of $\Psi^{(i-1)}_{\rho}$ is given by
\begin{align}
\max_{  \left\lVert \Psi^{(i-1)}_{\rho} \right\lVert} p_{1}^{(i)}= \frac{1}{1+(F_{i}^{-})^{2}},
\end{align}
where $F_{i}^{-}$ is defined in Eq. \eqref{eq:F} and depends on $B_{i}$.

The success probability, on the other hand, is $p_{\mathrm{success}}^{(i)} = 1-p_{\mathrm{failure}}^{(i)}$. Maximizing $p_{1}^{(i)}$, a first inequality can be written as
\begin{align}
p_{\mathrm{success}}^{(i)}
&\geq 1- \left(\max_{  \left\lVert \Psi^{(i-1)}_{\rho} \right\lVert} p_{1}^{(i)} \right)(1-s_{i}^{2}).
%
\end{align}
The right-hand side of the last expression is a monotonically decreasing function of $|B_{i}|$. As a consequence, the success probability is minimized when $|B_{i}|$ is maximized. As can be readily verified, $\max|B_{i}| = 1/N_{t}$ because the other terms have values in the interval $|w_{\rho}|,|u_{i}|,|e^{iEt}| \in [0,1]$.
%
%
Using this result, the success probability after $N_{t}+1$ applications of operator $\hat{B}_{i}$, as required by the algorithm and denoted by $P_{\mathrm{success}}$, is bounded by
\begin{align}
P_{\mathrm{success}} 
&\geq  \left[ \frac{1+ \frac{1}{4N_{t}^{2}} - \frac{1}{2N_{t}}\sqrt{1+\frac{1}{4N_{t}^{2}}}}{1+ \frac{1}{4N_{t}^{2}} + \frac{1}{2N_{t}}\sqrt{1+\frac{1}{4N_{t}^{2}}}} \right]^{N_{t}+1} .
\end{align}
For a large number of iterations, this becomes
\begin{align}
\label{eq:prob_success}
P_{\mathrm{success}} &\geq  \frac{1}{e} \left[ 1 - \frac{1}{N_{t}} \right]  + O(N_{t}^{-2}),
\end{align}
where $e \approx 2.7183$ is Euler's number. 

The success probability $P_{\mathrm{success}}$ relates to the implementation of the non-unitary operations that filter the trial function. As a consequence, the mean number of realizations required to implement the filtering part of the algorithm is approximately  $  e $. This is low enough that the efficiency of the global algorithm, including the time evolution, is not deteriorated by the nondeterministic nature of the spectral filtering method. However, the total success probability of the algorithm is
\begin{align}
P_{\mathrm{total}} &= P_{\rho}P_{\mathrm{success}}, \\
& = \frac{\langle \psi_{\rho}| \psi_{\rho}\rangle}{N^{2}e} + O(N_{t}^{-1}).
\end{align} 
Consequently, the total success probability of the whole procedure depends on both the probability of being in the eigenstate $P_{\rho}$ and the probability of success of the non-unitary operation. The total success probability and the accuracy of the method are analyzed in more detail in the next section.  

\section{Accuracy, complexity analysis and resource requirements}
\label{sec:complexity_an}

In this section, some properties of the filtering technique are studied quantitatively and compared to the ALT. In particular, estimates for the accuracy and for the resource requirements are obtained in terms of the filter properties. 

\subsection{Accuracy of the filtering method}

The accuracy of the filtering method is evaluated in Appendix \ref{app:accuracy}. If the suppression of the filter is large enough and if the trial function is real, the error is bounded by
\begin{align}
\label{eq:err_est_text}
\epsilon \leq \max \left[ CN_{t}^{-q}, \epsilon_{\mathrm{f}} \right].
\end{align} 
The first argument on the right-hand side of Eq. \eqref{eq:err_est_text} is the error of the time evolution discretization scheme, which is $O((\Delta t)^{q})$. The value of the constant $C$ depends on the Hamiltonian considered while $q$ depends on the order of accuracy of the time evolution and on the smoothness of the Hamiltonian \cite{1751-8121-43-6-065203}. The second argument defined by
\begin{align}
\epsilon_{\mathrm{f}} := \frac{S^{2}(1-A)}{A|L(0)|^{2}},
\end{align}
is the error of the filtering procedure, where $S$ is the minimum suppression of the filter (defined in Eq. \eqref{eq:supp_fact}), $A:=|\langle \varphi_{\rho}| \Psi_{\mathrm{trial}}(0) \rangle|^{2} $ is the overlap of the trial function with the desired eigenstate and $|L(0)| \in [0,1]$ is the coherent gain of the filter. Typical values for the coherent gain lie in the interval $0.3 \lesssim |L(0)| \lesssim 0.8$, with a value of  $|L(0)|=1$ for the rectangular window function \cite{harris1978use,heinzel2002spectrum}. Therefore,  the accuracy of the filtering is controlled by the filter properties (the suppression and the coherent gain) and by the initial trial state.  

The value of the suppression factor $S$ depends on the filter chosen. One popular choice is the Hann function \cite{Feit1982412}. However, there now exist high-performance filters where the side lobe suppression levels can reach up to -248~dB \cite{heinzel2002spectrum}, reducing the power spectral density by a factor of $\approx 10^{25}$ outside the range of interest and relative to the central frequency. As a comparison, a rectangular window function, as in the ALT, reaches -13.3~dB while the Hann window function has a suppression factor of -31.5~dB. Choosing a filter with a large suppression could potentially lead to a very accurate eigenstate estimation. 

\subsection{Performance of the method}

The high suppression factors described in the last section are possible only when the filter resolves the eigenenergies, when the energy width of the line shape is smaller than the energy interval between the wanted eigenenergy and its closest neighbor  as $\Delta E_{\mathrm{window}} < \Delta E_{\rho}:= \min|E_{\rho} - E_{\rho \pm 1}|$. Here, $\Delta E_{\mathrm{window}}$ is the bandwidth defined from the center of the filter $E_{\rho}$ to the minimum between the main peak and the first side lobe. The resolution of the filter is then related to the total time of the calculation as $\Delta E_{\mathrm{window}} =W\pi/T$, where $W \in \mathbb{R}^{+}$ is the line shape width. The rectangular filter has the best resolution with $W_{\mathrm{rectangular}}=1$. For other filters, with higher suppression strength, typical values are $2.0 \lesssim W \lesssim 6.0$. Therefore, having a higher suppression strength usually entails more time iterations to resolve the eigenstates.

The number of iterations $N_{t}$ required is determined by fixing $\Delta t$ and $T$ to obtain the desired bandwidth and resolution. First, the size of the time step $\Delta t$ should guarantee that the time-dependent evolution reaches convergence: this is achieved when the largest frequency component is resolved and when the numerical scheme is stable (as an example, for explicit time integration scheme in real space, a necessary condition for stability is the Courant-Friedrichs-Lewy condition \cite{leveque2002finite}). In addition, the value of $\Delta t$ should make sure that the error due to the time evolution is at least as small as the wanted accuracy. Finally, because the energy range of the filter is $[- \pi/\Delta t, \pi/\Delta t]$, yielding a bandwidth of $B:=2\pi/\Delta t$, the time step has to be set to give an energy range that accommodates all of the desired spectral components. Second, as mentioned earlier, the final time should be large enough such that eigenenergies are resolved. These considerations yield the following condition on the number of time steps:
\begin{align}
N_{t} > \max \left[\frac{ W B}{2(\Delta E_{\rho}) }, \left(\frac{C}{\epsilon} \right)^{\frac{1}{q}} \right].
\end{align}
This condition ensures that the filter resolves the eigenstate and that the desired accuracy is achieved.
Therefore, if a high resolution is required, the computation time is proportional to the width of the filter and inversely proportional to the energy interval between eigenvalues. For a given Hamiltonian, the number of time steps required is then $N_{t} = \mathrm{poly}(n)$, as long as the eigenstates are not exponentially close together, obeying $(\Delta E_{\rho})^{-1} = \mathrm{poly}(n)$. For many cases of interest, this condition is fulfilled and our initialization technique becomes exponentially faster than any classical algorithm. For well-separated eigenenergies, the number of time steps is rather determined by the accuracy of the time evolution. Then, the width of the filter is not critical to the performance of the method.

For quasigenerate states exponentially close to each other, when $\Delta E_{\rho} \sim 2^{-n}$, the quantum efficiency is absent because resolving these states necessitates an exponential number of time steps. One example is the  Coulomb-like system for which the eigenvalues display an accumulation point at $E=0$. Resolving the highly excited states in the vicinity of $E=0$ requires an exponential number of operations because of their quasi-degenerate nature. 

Also, this technique could cease to be efficient for many-body systems, in the limit of a large number of strongly correlated particles \cite{schuch2009computational}. In this case, the required accuracy decreases exponentially with the number of particles \cite{RevModPhys.71.1253}, demanding an exponential calculation time.  As a matter of fact, there exist classes of local Hamiltonian for which the eigenenergies are QMA-hard to obtain \cite{doi:10.1137/S0097539704445226} and therefore, are likely to involve an exponential number of operations. Nevertheless, although our method is not valid for all QMA-hard problems, it may still be useful in practical applications, such as quantum chemistry simulations \cite{kassal2011simulating}. These limitations are also shared by the ALT. 

The average number of operations $\bar{N}$ required to initialize the wave function using this quantum algorithm scales like $\bar{N} = e N_{t} \mathrm{poly}(n)/P_{\rho}$. This can be compared to the general states initialization methods described at the beginning, which scale like $O(2^{n})$. In the limit of a large number $n$ of qubits, the quantum spectral filtering method has an exponential gain of performance, if $P_{\rho}$ is not exponentially small. This probability is now discussed in more details. As long as  $\epsilon_{\mathrm{f}} \ll 1$, the probability can be estimated as
\begin{align}
P_{\rho} = \frac{A|L(0)|^{2}}{1+A|L(0)|^{2}} + O(\epsilon_{\mathrm{f}}),
\end{align} 
and thus, varies in the interval $P_{\rho}\in [0,1/2]$, according to the value of the overlap and the coherent gain. Similar to the accuracy, this probability depends on the coherent gain of the filter and on the trial state overlap. As mentioned earlier,  typical values of the coherent gain for most filters are $\sim 10^{-1}$, reducing the performance of the algorithm when compared to the rectangular filter. This probability is also controlled by the overlap, confirming that the trial state choice is highly important for the computational cost.

\subsection{Comparison with the ALT}

Using the estimates derived above, it is now possible to make a comparison with the ALT.
The computational complexity of the filtering algorithm is similar to the ALT: the number of operations in the ALT scales like $ \bar{N}^{(\mathrm{ALT})} = M \mathrm{poly}(n)/P_{\rho}^{(\mathrm{ALT})}$, where $M = N_{t}$, the number of iterations, determines the resolution and is similar to our number of time steps. However, the success probability is different and is given by $P_{\rho}^{(\mathrm{ALT})} = A + O(\epsilon_{\mathrm{f}})$ \cite{PhysRevLett.91.257902}. Combining the preceding results, it is possible to compare the resource requirements for both methods. We get 
\begin{align}
\frac{\bar{N}}{\bar{N}^{(\mathrm{ALT})}} < \frac{2e}{|L(0)|^{2}} \frac{N_{t}}{N_{t}^{(\mathrm{ALT})}}.
\end{align} 
For applications where the resolution is critical, this becomes $\bar{N}/\bar{N}^{(\mathrm{ALT})} < 2eW/|L(0)|^{2}$ and the performance is dictated by the coherent gain and the width of the filter. However, if the accuracy determines the number of time steps, we have $\bar{N}/\bar{N}^{(\mathrm{ALT})} < 2e/|L(0)|^{2}$ and only the coherent gain is important for the performance.

In terms of the accuracy, the comparison between the two methods yields
\begin{align}
\frac{\epsilon}{\epsilon^{(\mathrm{ALT})}} = \frac{S^{2}}{S^{2}_{\mathrm{rectangular}}|L(0)|^{2}},
\end{align}
where it is assumed that $CN_{t}^{-q} < \epsilon_{\mathrm{f}},\epsilon_{\mathrm{f}}^{(\mathrm{ALT)}}$. Therefore, the accuracy of our filtering technique can potentially be superior to the ALT.

However, there is clearly a tradeoff between accuracy and performance, that depends on the type of filter and the technique used. In particular, the ALT is more efficient by a constant factor that depends on the properties of the filter. On the other hand, our technique can potentially yield more accurate results if a larger suppression factor is considered.

Moreover, our technique does not require a large number of ancilla qubits, a useful property given that the number of qubits is limited to $\approx 14$ in actual quantum computer registers \cite{PhysRevLett.106.130506,ladd2010quantum,0034-4885-74-10-104401}. 
A typical quantum simulation in real or momentum space for a non-relativistic single particle system, for example, requires at least $\approx$ 256-512 lattice points in 1D to reach convergence, making for a quantum register having more than $\approx$ 8-9 qubits to store the wave function. For the ALT method, the register for the quantum Fourier transform requires approximately $\approx$ 6-7 additional qubits \cite{PhysRevLett.83.5162}. In addition, a certain number of qubits should be reserved for error correction and possibly, for additional work space. Therefore, using ALT, the resource requirement for this type of quantum simulation is at the limit and probably above the resources available. With our algorithm, only two ancilla qubits are necessary, for any energy resolution. This reduction of the number of qubits could allow for the simulation of a certain type of systems using actual quantum computers, whereas they would be out of reach using the ALT. This of course, assumes that the coherence time of the quantum device can accommodate the larger computational cost of our method.

\section{Specific example: the harmonic oscillator}
\label{sec:harm_osci}

In this section, some features of the filtering method are illustrated in a simple example: the quantum harmonic oscillator. The Hamiltonian, expressed in the oscillator natural units where the energy is in units of $\hbar \omega$, lengths are in units of $\sqrt{\hbar/(m\omega)}$, $\omega$ is the oscillator frequency and $m$ the particle mass, is given by
\begin{align}
\hat{H} = \frac{\hat{p}^{2}}{2} + \frac{1}{2}\hat{x}^{2}.
\end{align}
The ground state of this system is now generated on a classical computer by the filtering method combined with techniques borrowed from real space quantum simulation. In particular, the 1D time-dependent Schr\"{o}dinger equation is solved by using the pseudospectral second order split operator scheme described in Appendix \ref{app:td_sh}. The latter could be implemented efficiently on a quantum computer by using the quantum Fourier transform \cite{OPPROP:PR877,Strini2008}.

The filtering is applied to a trial state given by
\begin{align}
\Psi_{\mathrm{trial}}(0,x) =
\begin{cases}
\cos^{2} \left( \frac{\pi x}{2\ell} \right), & x\in[-\ell,\ell]\\
0,& \mathrm{else}
\end{cases} 
\end{align}
where $\ell$ is the wave function width. This trial state can be generated efficiently with $\mathrm{poly}(n)$ quantum gates using the technique given in Refs. \cite{Zalka08011998,kaye2004quantum,grover2002creating} because its square can be integrated analytically. 

\begin{figure}
\includegraphics[width=0.5\textwidth]{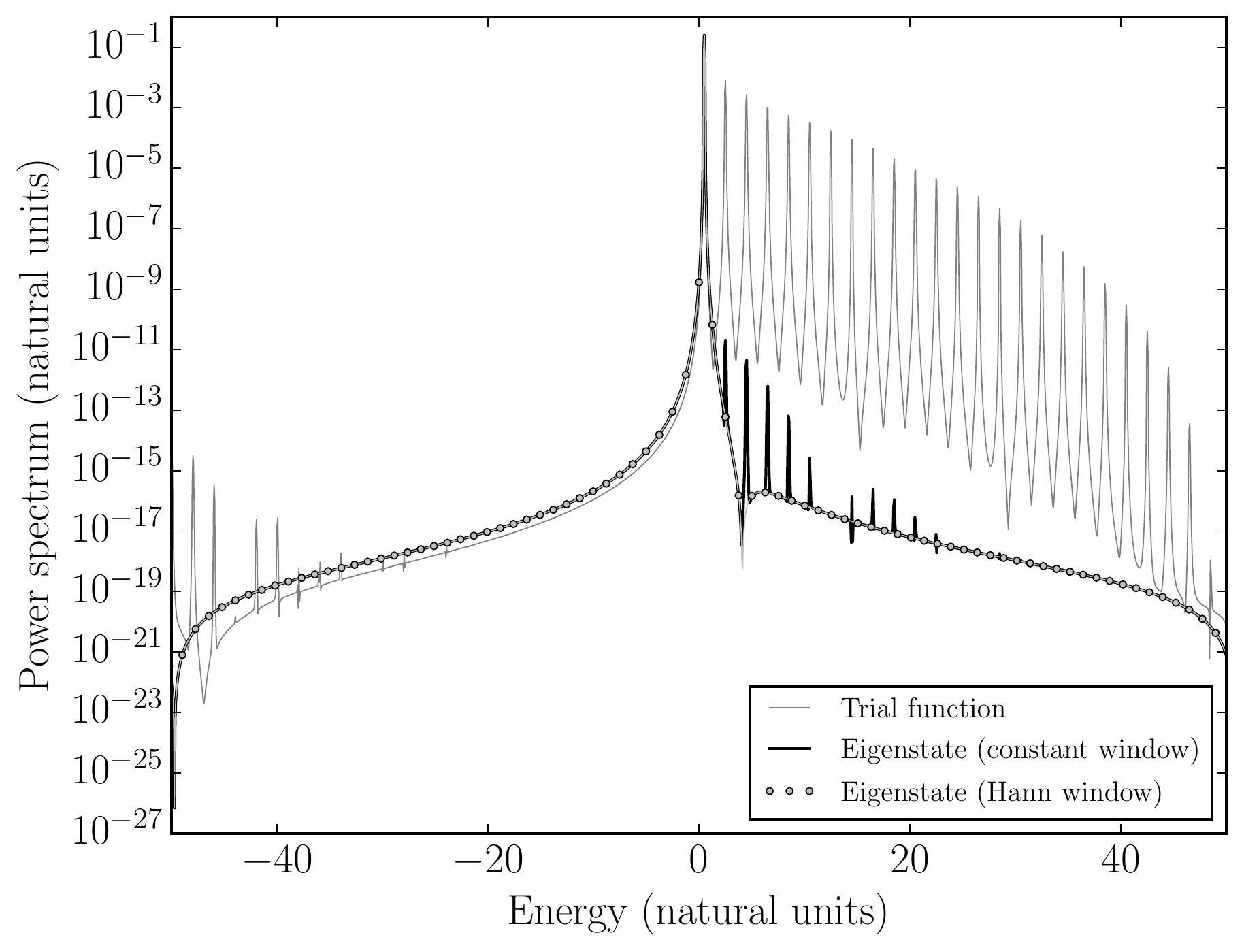}
\caption{Power spectra for the ground state generated from the filtering method using the constant or Hann window function.  The spectrum is also given for the trial state, prior to filtering.}
\label{fig:power_spec}
\end{figure}

The domain has a length of $L=40$ (all quantities are in the quantum oscillator natural units) and is discretized by 1024 lattice points. The trial state width is set to $\ell = 10$ and is evolved to a final time of $T=100$ using $N_{t} = 8192$ time steps. The power spectrum of the trial function is computed by using Eq. \eqref{eq:auto_corr} and the results are displayed in Fig. \ref{fig:power_spec}. It can be verified that the energy of the highest peaks in the power  spectrum are positioned at energies 
\begin{align}
E = E_{m} = m + \frac{1}{2}, \;\; m =1,3,5,\cdots,
\end{align}
in agreement with the analytical solution of the harmonic oscillator. Components with $m$ even do not appear in the spectrum because they are antisymmetric while the trial wave function is symmetric. 

The trial state is then filtered using Eq. \eqref{eq:eigen_state}. The simulation parameters are the same as above. Two window functions are considered: a rectangular window as in the ALT and the Hann window, given by
\begin{align}
w(t) = \frac{1}{2}\left[ 1 - \cos\left(\frac{2\pi t}{T} \right) \right].
\end{align} 
The resulting power spectrum of the filtered ground state (with $E_{0} = 1/2$) is also displayed in Fig. \ref{fig:power_spec}. For both window types, the power spectrum is sharply peaked around the ground state energy. However, the Hann window function suppresses the unwanted modes by up to five orders of magnitude more than the rectangular window function, conferring a distinct advantage of our apodization technique because it yields more accurate states. This is confirmed by computing the numerical error $\epsilon = \left\lVert \psi_{\mathrm{filtered}} - \psi_{\mathrm{exact}} \right\lVert$. We find that $\epsilon_{\mathrm{Hann}} \approx 2.42 \times 10^{-8}$ while $\epsilon^{(\mathrm{ALT})} = \epsilon_{\mathrm{rect}} \approx 1.77 \times 10^{-5}$.

The calculation considered here could be performed on a 12 qubit register: 10 qubits are used to store the wave function and two ancilla qubits are utilized in the initialization phase, as discussed previously. In comparison, for the same energy resolution, the ALT would require approximately 23 qubits: 10 qubits for the wave function and $\approx 13$ qubits for the quantum Fourier transform. Given that the state-of-the-art quantum registers have $\approx 14$ qubits \cite{PhysRevLett.106.130506}, reducing the number of ancilla qubits is still an important issue for the success of quantum simulations. Therefore, our quantum filtering technique is an interesting alternative from this standpoint.   

The success probability, on the other hand, favors the ALT. The overlap of the trial state with the desired state yields a success probability of $P^{(\mathrm{ALT})}_{\mathrm{total}} \approx 0.45$. For our filtering technique, this probability is evaluated explicitly using Eqs. \eqref{eq:p1_app}, \eqref{eq:p_failure} and the expression of $P_{\rho}$. We find that $P_{\mathrm{total}} \approx 0.061$. Therefore, our filtering scheme demands more computation time than for ALT, demonstrating the tradeoff between accuracy, computation time and the number of qubits required.   

Interestingly, the accuracy of our filtering scheme is approximately the same as for ALT using a number of time steps given by $N_{t} = 1600$, using the same value for other parameters. In this case, the accuracy is reduced to $\epsilon_{\mathrm{Hann}} \approx 1.66 \times 10^{-5}$ while the change in the probability of success is negligible. Therefore, one obtains that $\frac{\bar{N}}{\bar{N}^{(\mathrm{ALT})}} \approx 1.44$, showing that the performance overhead can be relatively unimportant in some cases for a fixed accuracy, although the number of required qubits is reduced.

\section{Conclusion}
\label{sec:conclusion}

In conclusion, the algorithm we are presenting allows for an efficient initialization of a quantum register to a state with a specified energy range under some specific conditions: the overlap of the trial state with the eigenstate and the energy difference between eigenstates should not be exponentially small. Our algorithm improves upon the ALT on two main aspects: the inclusion of a window function allows for more accurate eigenstates (the accuracy depends on the choice of filter) and the number of ancilla qubit is limited to two, for any energy resolution. In the ALT, this energy resolution dictates the number of qubits. However, our technique generally requires more quantum operations than the ALT and thus, there is a tradeoff between accuracy, computation time and resource requirements. These features were demonstrated in the simple example of the quantum harmonic potential system. 

Such a procedure is important for dynamical quantum simulations on digital quantum computers where the goal is to simulate the time-dependent behavior of the wave function and to determine scattering process probabilities where the initial state is one specific eigenstate of some static Hamiltonian. These are important in many fields such as atomic, molecular, optical and condensed matter physics. 

Furthermore, our algorithm can be applied to many physical systems in any basis (real space, momentum space, etc) as long as an efficient time evolution algorithm is available and the conditions given above are fulfilled. Other explicit implementations for the solution of quantum equations will be given elsewhere.  

Finally, it may be interesting to apply the novel concept of qubitization to our filtering method. This new technique promises optimal query complexity for the computation of a large class of unitary and non-unitary operators \cite{low2016hamiltonian}. This will be investigated further in the future.  

\appendix

\section{Efficient measurement of the energy spectrum}
\label{app:en_spec}

The energy spectrum of a given time-independent Hamiltonian can be obtained from the time evolution of a trial function by calculating the Fourier transform of the autocorrelation function $ \langle \Psi_{\mathrm{trial}}(0)|\Psi_{\mathrm{trial}}(t) \rangle$, as in Eq. \eqref{eq:auto_corr}. Again, this is an adaptation of the Feit-Fleck method \cite{Feit1982412} to quantum computing and was already discussed in the context of DQC1 \cite{PhysRevLett.81.5672} and for the simulation of physical systems \cite{PhysRevA.65.042323}. We give here a simple formulation of this algorithm. We note that these approaches are similar in spirit to the one-qubit implementation of the quantum Fourier transform in Schor's algorithm \cite{PhysRevLett.76.3228}.

\begin{figure}
\includegraphics[scale=1.0]{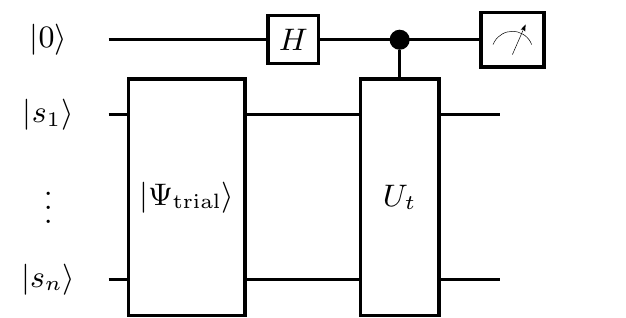}
\caption{Circuit diagram for the quantum implementation of the measurement of the correlation function. The gate $U_{t}$ advances the trial solution by $t$. The gate $H$ is a Hadamard gate. The gate $|\Psi_{\rm trial}\rangle$ initializes the quantum register to an arbitrary state. Finally, the quantity $\langle \sigma_{x,y} \rangle$ are measured on the ancilla qubit. }
\label{fig:autocorr}
\end{figure}

An ancilla qubit is added in the state $|c\rangle = |0\rangle$ to double the register and a random trial function is implemented on the other qubits, setting the quantum register in the state $|0\rangle \otimes |\Psi_{\mathrm{trial}}(0)\rangle$. Then, the following mappings are performed:
\begin{align}
& |0\rangle \otimes |\Psi_{\mathrm{trial}}(0)\rangle \\
&\mapsto \frac{1}{\sqrt{2}} \left[|0\rangle \otimes |\Psi_{\mathrm{trial}}(0)\rangle + |1\rangle \otimes |\Psi_{\mathrm{trial}}(0)\rangle \right] ,\\
&\mapsto \frac{1}{\sqrt{2}} \left[|0\rangle \otimes |\Psi_{\mathrm{trial}}(0)\rangle + |1\rangle \otimes |\Psi_{\mathrm{trial}}(t)\rangle \right] .
\end{align}
These mappings are easily implemented using a Hadamard gate, doubling the register to store the initial trial state, at $t=0$. The second mapping is a controlled evolution operator, which evolves the trial state to time $t$. The resulting circuit diagram is displayed in Fig. \ref{fig:autocorr}. 

The last step of the quantum algorithm is the measurement of $\langle \sigma_{x,y} \otimes \mathbb{I}\rangle$, where $\sigma_{x,y}$ are Pauli matrices, on the ancilla qubit (here, the identity is in the subspace of qubits that stores the wave function). It can then be demonstrated that
\begin{align}
\langle \sigma_{x} \otimes \mathbb{I}\rangle &= 2 \mathrm{Re} \langle \Psi_{\mathrm{trial}}(0)|\Psi_{\mathrm{trial}}(t) \rangle ,\\
\langle \sigma_{y} \otimes \mathbb{I}\rangle &= 2 \mathrm{Im} \langle \Psi_{\mathrm{trial}}(0)|\Psi_{\mathrm{trial}}(t) \rangle .
\end{align} 
Therefore, from these measurements, it is possible to construct the autocorrelation function. This algorithm is efficient if the time evolution operator $U$ is also efficient. However, it requires the use of classical computation to evaluate the Fourier transform in Eq. \eqref{eq:auto_corr} to obtain the spectrum. Then, the eigenenergies can be read off from the spectrum.

\section{Filtering method}
\label{app:filtering}

In this appendix, we review some standard results for the filtering method relevant to our quantum algorithm. In particular, the spectrum of the filtered signal is evaluated to reveal the effect of the window function.

Any general time-dependent state, using an eigendecomposition, can be written as 
\begin{align}
\label{eq:eigen_decomp}
|\Psi_{\mathrm{trial}}(t) \rangle = \sum_{m,j_{m}} a_{m,j_{m}} e^{-iE_{m}t} |\varphi_{m,j_{m}} \rangle,
\end{align}
where $E_{m}$ is the discrete eigenenergy, $a_{m,j_{m}}$ is the spectral coefficient of a given mode, $j_{m}$ is an index over degenerate states while $|\varphi_{m,j_{m}} \rangle$ are the eigenvectors. On a digital quantum computer, there is no continuum spectrum because the computer simulates a compact system. As a consequence, the spectrum is always discrete. Multiplying by $w_{\rho}(t)e^{iE_{\rho}t}/T$ and integrating on time, we get
\begin{align}
\label{eq:init_filtered}
|\Psi_{\rho} \rangle = \frac{1}{T} \int_{0}^{T} dt w_{\rho}(t) e^{iE_{\rho}t} |\Psi_{\mathrm{trial}}(t)\rangle,
\end{align}
where $|\Psi_{\rho} \rangle$ is the state filtered by the window function $w_{\rho}(t)$ and $E_{\rho}$, as shown below, is the central frequency of the filter. 

To evaluate the effect of the window function $w_{\rho}$, we compute the spectrum of the state $|\Psi_{\rho} \rangle$. The power spectrum of this state can be obtained from the Fourier transform of the autocorrelation function as
\begin{align}
\label{eq:auto_corr}
C(E) = \frac{1}{T}\int_{0}^{T} dt w(t) e^{iEt} \langle \Psi_{\rho}(0)|\Psi_{\rho}(t) \rangle,
\end{align}  
where $w(t)$ is a window function allowing for a finite time integration. Using Eq. \eqref{eq:init_filtered}, the last equation can be written as
\begin{align}
C(E) &= \frac{1}{T^{3}}\int_{0}^{T} dtdt_{1}dt_{2} w(t) e^{iEt + iE_{\rho}(t_{2}-t_{1})} w_{\rho}(t_{1})w_{\rho}(t_{2}) \nonumber \\
& \times \langle \Psi_{\mathrm{trial}}(t_{1})| \hat{U}(0,T) |\Psi_{\mathrm{trial}}(t_{2}) \rangle.
\end{align}
Decomposing the trial state using the eigen-decomposition in Eq. \eqref{eq:eigen_decomp} and simplifying, we get
\begin{align}
\label{eq:spec_coeff}
C(E) &= \sum_{m,j_{m}} |a_{m,j_{m}}|^{2} |L_{\rho}(E_{\rho} - E_{m})|^{2} L(E-E_{m}),
\end{align}
where $L_{\rho}$ is the line shape given by
\begin{align}
\label{eq:line_shape_app}
L_{\rho}(E_{\rho}-E_{m}) = \frac{1}{T} \int_{0}^{T}dt e^{i(E_{\rho}-E_{m})t} w_{\rho}(t),
\end{align}
and $L$ is the line shape associated to $w(t)$. Therefore, the power spectrum consists in Dirac-delta-like peaks $L$, representing energy modes centered on $E_{m}$, weighted by $\sum_{j_{m}} |a_{m,j_{m}}|^{2} |L_{\rho}(E_{\rho} - E_{m})|^{2}$. Therefore, the spectral content of the filtered signal is determined both by the trial function through $a_{n,j}$ and by the window function $w_{\rho}$. For carefully designed filters, the suppression of unwanted modes, accomplished by $L_{\rho}$, can be exponentially large.

\section{Expressions resulting from the singular value decomposition}
\label{app:svd}

The operator $\hat{B}_{i}$ is re-expressed by using a singular value decomposition as $\hat{B}_{i} = U_{i}\Sigma_{i}V_{i}^{\dagger}$. These matrices are written explicitly as
\begin{align}
U_{i} &= 
\begin{bmatrix}
\frac{F^{+}}{\sqrt{(F_{i}^{+})^{2} + (G_{i}^{+})^{2}}} &
\frac{F^{-}}{\sqrt{(F_{i}^{-})^{2} + (G_{i}^{-})^{2}}} \\
\frac{G^{+}}{\sqrt{(F_{i}^{+})^{2} + (G_{i}^{+})^{2}}} &
\frac{G^{-}}{\sqrt{(F_{i}^{-})^{2} + (G_{i}^{-})^{2}}} 
\end{bmatrix} \\
\label{eq:V_svd}
V_{i}^{\dagger} &= 
\begin{bmatrix}
\frac{F^{+}}{\sqrt{(F_{i}^{+})^{2} + 1}} &
\frac{1}{\sqrt{(F_{i}^{+})^{2} + 1}} \\
\frac{F^{-}}{\sqrt{(F_{i}^{-})^{2} + 1}} &
\frac{1}{\sqrt{(F_{i}^{-})^{2} + 1}}
\end{bmatrix} 
\end{align} 
where we defined
\begin{align}
\label{eq:F}
F_{i}^{\pm} &:= \frac{B_{i}}{2} \pm  \sqrt{1+\frac{B_{i}^{2}}{4}}, \\
\label{eq:G}
G_{i}^{\pm} &:= 1 + \frac{B^{2}_{i}}{2} \pm |B_{i}| \sqrt{1+\frac{B_{i}^{2}}{4}}.
\end{align}

\section{Maximum bound on $p_{1}^{(i)}$}
\label{app:bound_p}

This appendix is devoted to finding the maximum value of 
\begin{align}
p_{1}^{(i)} := \frac{\langle \psi_{1}^{(i)}|\psi_{1}^{(i)}\rangle}{1+\langle \Psi_{\rho}^{(i-1)}|\Psi_{\rho}^{(i-1)} \rangle}.
\end{align}
Using Eq. \eqref{eq:V_svd}, the last equation is given by
\begin{align}
\label{eq:p1_app}
p_{1}^{(i)} &= \frac{|(V_{i}^{\dagger})_{21}|^{2} + |(V_{i}^{\dagger})_{22}|^{2}  \langle \Psi_{\rho}^{(i-1)}|\Psi_{\rho}^{(i-1)} \rangle}{1+\langle \Psi_{\rho}^{(i-1)}|\Psi_{\rho}^{(i-1)} \rangle}. 
\end{align}
This last function is a monotonically increasing function of the norm $\langle \Psi_{\rho}^{(i-1)}|\Psi_{\rho}^{(i-1)} \rangle$ and therefore, is maximized when the norm is large. Therefore, it yields
\begin{align}
\max_{  \left\lVert \Psi^{(i-1)}_{\rho} \right\lVert} p_{1}^{(i)}  &=  |(V_{i}^{\dagger})_{22}|^{2}  = 
\frac{1}{(F_{i}^{-})^{2} + 1}.
\end{align}

\section{Accuracy estimate of the filtering method}
\label{app:accuracy}

In this appendix, a bound on the error of the filtering method is derived. We assume there is no degenerate states. The error is defined as
\begin{align}
\epsilon = \left\lVert \varphi_{\rho} - \Psi_{\rho} \right\lVert,
\end{align}
where $|\varphi_{\rho}\rangle$ is the wanted eigenstate while $|\Psi_{\rho}\rangle$ is an approximation of the eigenstate obtained from the spectral filtering method. Here, both states are normalized to 1. Using Eqs. \eqref{eq:eigen_decomp} and \eqref{eq:init_filtered}, the error can be written as
\begin{align}
\epsilon = 1+ \frac{|a_{\rho}|^{2}|L(0)|^{2}}{N_{\rho}^{2}} - 2 \frac{|a_{\rho}||L(0)|}{N_{\rho}} \cos(\theta) + \frac{A_{\rho}}{N_{\rho}^{2}},
\end{align} 
where $\theta$ is the phase of the amplitude $a_{\rho}$, where we defined
\begin{align}
A_{\rho}:= \sum_{m \neq \rho} |a_{m}|^{2}|L(E_{\rho}-E_{m})|^{2},
\end{align}
and where the normalization of $|\Psi_{\rho}\rangle$ is given by
\begin{align}
N_{\rho}^{2} = |a_{\rho}|^{2}|L(0)|^{2} + A_{\rho}.
\end{align}
When the suppression of the filter is large enough, we have that $A_{\rho} \ll |a_{\rho}|^{2}|L(0)|^{2}$, allowing us to expand the expression of the error, yielding
\begin{align}
\epsilon = 2 - 2\cos(\theta) + \frac{A_{\rho}}{|a_{\rho}|^{2}|L(0)|^{2}}\cos(\theta) + O\left(\frac{A_{\rho}^{2}}{|a_{\rho}|^{4}|L(0)|^{4}}\right).
\end{align}
This expression shows that the phase should be $\theta = 0$ to minimize the error, implying that the trial state should be real. Then, we introduce a minimal suppression factor defined as
\begin{align}
\label{eq:supp_fact}
S:= \max_{m \neq \rho}|L(E_{\rho}-E_{m})|,
\end{align}
which allows writing the inequality
\begin{align}
A_{\rho} \leq S^{2} \sum_{m \neq \rho} |a_{m}|^{2} = S^{2}(1-|a_{\rho}|^{2}).
\end{align}
Then, we have the overlap $A:=|\langle \varphi_{\rho}| \Psi_{\mathrm{trial}}(0) \rangle|^{2} = |a_{\rho}|^{2}$. Combining all of these results yields an estimate of the error of the filtering method given by
\begin{align}
\label{eq:err_est}
\epsilon \leq \frac{S^{2} (1-A)}{A|L(0)|^{2}}.
\end{align} 

\section{Solving the time-dependent Schr\"{o}dinger equation in real space}
\label{app:td_sh}

The time-dependent Schr\"{o}dinger equation in real space representation is 
\begin{align}
i\partial_{t} \psi(x,t) = \left[ \frac{\hat{p}^{2}}{2} + V(x) \right] \psi(x,t) ,
\end{align}
where $\psi(x,t)$ is the wave function, $\hat{p} = -i d/dx$ is the momentum operator and $V(x)$ is a scalar potential. The solution of this equation is given by
\begin{align}
\psi(x,t_{i}+\delta t) = e^{-i \delta t \left[ \frac{\hat{p}^{2}}{2} + V(x) \right]} \psi(x,t_{i}),
\end{align} 
where $t_{i}$ is the initial time and $\delta t$ is a small time increment. The evolution operator can be approximated by a second order split operator method as
\begin{align}
\psi(x,t_{i}+\delta t) &=e^{-i \frac{\delta t}{2} V(x)} e^{-i \delta t \frac{\hat{p}^{2}}{2}} e^{-i \frac{\delta t}{2} V(x)}\psi(x,t_{i}) \nonumber \\
& + O(\delta t^{3}).
\end{align} 
The momentum operator is diagonal in Fourier space. Taking advantage of this fact, the pseudospectral method consists in taking the Fourier transform of the wave function and write the evolution as \cite{doi:10.1021/j100319a003,tannor2007introduction}
\begin{align}
\psi(x,t_{i}+\delta t) &=e^{-i \frac{\delta t}{2} V(x)} \nonumber \\
&\times  \mathrm{FT}^{-1}\left\{ e^{-i \delta t \frac{p^{2}}{2}} \mathrm{FT} \left\{e^{-i \frac{\delta t}{2} V(x)}\psi(x,t_{i}) \right\}\right\} ,
\end{align} 
where $\mathrm{FT}(\cdot)$ denotes the Fourier transform operator. This numerical scheme can be implemented easily on a classical computer and can be implemented efficiently on a quantum computer \cite{OPPROP:PR877,Strini2008}.

\begin{acknowledgments}
The authors would like to acknowledge interesting discussions with R. Somma. Also, the authors are grateful to J. Dumont for carefully reviewing earlier versions of the manuscript. Finally, we thank anonymous referees for insightful comments. 
\end{acknowledgments}

\bibliographystyle{apsrev4-1}
\bibliography{bibliography}

\end{document}